# Ageing Effect of $Sb_2Te_3$ thin films

P. Arun, Pankaj Tyagi, A. G. Vedeshwar and Vinod Kumar Paliwal

Department of Physics & Astrophysics

University of Delhi, Delhi 110 007

INDIA.

## ABSTRACT

Post deposition variation in film resistance of $Sb_2Te_3$ films deposited on glass substrates at room temperature and an elevated temperature ($110^oC$) are investigated. The resistance of films grown at room temperature shows a non-linear decrease with time which is thickness dependent as opposed to the increasing resistance of film grown at elevated temperature. The decreasing resistance with time can be attributed to the transformation of an amorphous phase of the as-grown film to a micro-crystalline phase as revealed by X-ray diffraction. The increasing resistance was found to be due to the surface oxidation ($Sb_2O_3$) and a diffusion as a function of time. However, the underneath layer of $Sb_2Te_3$ below the top $Sb_2O_3$ layer remains amorphous even after two years from the date of fabrication.

# 1. Introduction

Post deposition variation in the physical properties of thin films as a function of time is called ageing. Such post deposition variation has been reported in various films, like silver, copper [1] and $CdSe_XTe_{1-X}$ [2, 3] etc. In these studies the post deposition film resistance was found to increase with time and then saturate. Many theories have been put forward to explain this increase in film resistance. Agglomeration of the islands within the film, where coalescence is the force driving the agglomeration process, has been considered the cause of increase in film resistance. The increase in resistance was shown to be directly proportional to time ($lnR(t)/R(0)$ $\alpha$ $lnt$). The conduction mechanism across the voids between the islands/ grains is considered to be either due to quantum tunnelling or electron emission [4]. Fehlner [5], Erhlich [6] and Despande [7] assumed the oxidation of the metal islands' surface as the cause for the increasing resistance. This causes increase in the average inter-island spacing and a change in work function which leads to a decrease in tunnelling probability, thus leading to an increase in the films resistance with time. Morris [8] however explained an increase in resistance due to a reduced electron emission due to a decrease in film temperature. In short, the literature is full of examples where the resistance of the film increases with time. The properties of $Sb_2Te_3$ films also show a variation with time, immediately after deposition. However, it is seen that in films grown on glass substrates at room temperature, the resistance falls with time. Films of different materials seem to have unique process leading to ageing. Various peculiarities seen during the ageing of $Sb_2Te_3$ films have been reported in this manuscript. Therefore, it prompted us to study ageing effect of $Sb_2Te_3$ films in detail as reported here.

# 2. Experimental

Thin films of $Sb_2Te_3$ were grown on glass substrates kept at room temperature, using thermal evaporation method. $Sb_2Te_3$ ingot of high purity (99.99%) supplied by Aldrich (USA) were used as the starting material. The crushed ingot were evaporated from molybdenum boat at a vacuum better than $10^{-6}$ Torr. The film thickness was measured using Dektek IIA surface profiler which uses the method of a mechanical stylus movement on the surface. The movement of the stylus across the edge of the film determines the step height or the films thickness. The film thickness was found to be uniform over the area 5cm x 5cm with an error of 3% at the edges.

Before the glass substrates were placed in the chamber, indium contacts were grown on them. A strip of $Sb_2Te_3$ film of dimensions 2.3cm x 1.65mm could be fabricated on the pre-grown indium contacts using a mask. The I-V characteristics of the films were measured by four probe method. It was found to be linear between $25\mu V$-24V, showing the ohmic nature of indium contacts. The variation in films' resistance with time were measured by a digital multimeter. The variation in resistance of films grown at room temperature were done both in vacuum and at atmospheric pressure. However, variation in resistances of films grown on heated substrates was studied only at atmospheric pressure. The structural and compositional analysis of these films were done using Phillips PW1840 X-ray diffractometer and Shimadzu ESCA750 (Electron Spectroscopy for Chemical Analysis). The films were found to be stoichiometrically uniform over the area 5cm x 5cm as determined by ESCA carried out in various regions of the film.

## 3. Results and Discussion

We have observed a quite different nature of ageing effect for films grown at room temperature and elevated temperatures. The resistance of as grown films decreases with time non-linearly and shows film thickness dependence as shown in fig 1. However, the resistance of films grown at elevated temperature ($110^oC$) shows an initial linear increase and then a parabolic increase with time. We will discuss it separately in contrast with that of at room temperature. Coming to the case of films grown at room temperature, the decrease in resistance with time is irrespective of whether the film was kept in vacuum or in air at ambient. The decrease in resistance with time immediately rules out the possibility of any oxidation even when film is exposed to air which has been confirmed by ESCA. Almost all earlier studies [4-7] invoke the idea of oxidation in the film to explain the increasing resistance with time. This is quite probable in case of metal films and some compound's films containing few unreacted chemical species. However, the decreasing resistance with time is hardly reported in literature.

The thickness dependent behaviour of the decreasing film resistance with time is illustrated in figure 1. We can notice few things clearly. First of all, the sharp decrease in resistance commences after some time of initial slow decrease which is thickness dependent. Thinner the film, earlier is the commencement of the sharp decrease. Secondly, the duration of sharp decrease also shows thickness dependence. This duration increases with film thickness. Thirdly, after the sharp decrease the films resume again a slowly decreasing behaviour and saturate after a long time. The saturation time shows a linear dependence with time as displayed in fig 2.

The film resistance or in other words resistivity measured either immediately after deposition or after saturation shows similar systematic behaviour with film thickness as shown in figure 3 and 4 respectively. We have carried out the structural analysis of the films as a function of time by X-ray diffraction to get any clue regarding the decreasing resistance of the films. <u>A film of thickness 380nm was selected, since the larger the thickness, slower is the variation in resistance with time</u>. We have shown the results of X-ray diffraction as a function of time in figure 5.

All the as grown films were amorphous without exception as revealed by diffractogram (a) in fig 5. Till the diffractogram (d) there is no much direct revelation of structural changes. The micro-cyrstalline nature of the film can clearly be realised by the appearance of a broad peak in diffractogram (d) of fig 5, which was taken after the film resistance saturated. However, a close look at the diffractograms would indicate the evolution of micro-crystalline phase from an amorphous phase. Therefore, we believe, the decreasing film resistance with time is mainly due to the evolution of micro-crystallinity from an amorphous phase. However, the onset and end point of the sharp decrease in the film resistance of fig 1 could not be identified explicitly in the structural analysis.

The evolution of micro-crystalline phase starts from the relatively unstable amorphous phase through the atomic re-arrangements correcting the 'wrong bonds', normally existing even in stoichiometric amorphous phase, at a nucleation centre. Considering the various bond strengths, i.e. Sb-Sb (3.11eV), Te-Te (2.67eV) and Sb-Te (2.88eV) in $Sb_2Te_3$ [9], we can see that the difference between the average bond strengths of (Sb-Sb, Te-Te) and 2(Sb-Te) is nearly 20meV, comparable to the thermal energy at room temperature. Therefore, the atomic arrangement can take place by the thermal vibration of atoms resulting in 'right bonds' leading to a final micro-crystalline phase. The initial time lag for the commencement of sharp decrease in R(t) showing a linear thickness dependence could be assumed to be the initialisation time for the re-arrangement process. Further, the time duration for either the linear sharp fall of R(t) or the saturation also shows a linear thickness dependence. This could be explained if we assume the amount of 'wrong bonds' to be proportional to film thickness which is quite reasonable. This kind of atomic re-arrangement can not be followed by a simple X-ray diffraction analysis. However, the change in film resistance could easily be thought as due to a better connectivity for electron conduction resulting from atomic re-arrangements. Returning to figure 3 and 4, it should be noted that the nature of resistivity's variation with thickness is identical. It is as if there is a significant background initially, which diminishes with time. The resistivity is explained by Matthiessen's additive rule [10], which states the contributions due to thermal resistivity, from point defects, dislocations etc add up linearly

$$\rho(d) = \rho_T(d) + \rho_p(d) + \rho_d(d) + .......... \quad (1)$$

Since the change in resistance is attributed to increasing crystallinity, it may be assumed that the decrease in resistivity is due to the decrease in contributions from those of point defects and dislocations. However, since these terms, themselves are thickness dependent, a quantitative estimate is difficult. The variation in resistivity with film thickness is inversely proportional to the film thickness and is in accordance to Mayadas equation [11].

The resistance was found to be increasing with time linearly initially and then parabolically for films grown at an elevated temperature of $110^{o}C$ and exposed to air immediately after deposition as shown in fig 6. The X-ray diffractograms taken immediately after fabrication, showed these samples were amorphous in nature.

The increasing film resistance of a quite thick film can be suspected as due to oxidation and diffusion in the film. Therefore, we have carried out ESCA analysis, figure 7, on these films which shows the surface oxidation. The oxidation can be detected by the intensity ratio of antimony's $3d_{3/2}$ and $3d_{5/2}$ peaks which is equal to 0.66 for pure antimony. Since Sb $3d_{5/2}$ and O 1s peaks merge due to identical binding energies, the intensity of the undeconvoluted Sb $3d_{5/2}$ peak increases so that the ratio will decrease from 0.66, as explained by Morgan et al [12]. We have observed the ratio to due to be 0.44 indicating oxidation. The absence of Te peaks in the survey scan indicates the complete oxidation at the surface. Since ESCA is a surface technique limited to about 5nm depth [13], the estimate of the extent of oxidation and diffusion is not possible by this technique. The oxidation and diffusion of oxide in the film has been explained by Sze [14]. The thickness of oxide layer at the surface increases linearly with time initially. The changing film resistance due to the initial oxidation is given by

$$R(t) = \frac{\rho l / w d}{1 - Bt / Ad} \qquad (1)$$

where B/A is called the linear rate constant, a measure of the rate of oxidation. We get the linear rate constant as $50.16 \times 10^{-3}\,\mu m/hr$ by fitting eqn 1 to the initial linear increase of film resistance with time of fig 6. After the initial oxidation, the surface oxide layer acts as a protective layer preventing further oxidation. However, the film resistance continues to increase due to the diffusion of oxide into the depth of the film. The diffused layer thickness increases parabolically due to which the film resistance varies according to

$$R(t) = \frac{rl/wd}{1 - \frac{\sqrt{Bt}}{d}} \qquad (2)$$

where B is called parabolic rate constant. Eqn (2) fits very well to the experimental data as shown in fig 6 by the continuous curve yielding B =10.1 x $10^{-6} \mu m^2$/hr. We estimate the initial oxide layer thickness of about 25nm and a total diffused layer thickness of about 61nm in a 380nm thick film using the above analysis.

The resistivity of films grown on heated substrates were found to be only 10-55% of that grown at room temperature, with thinner films showing larger variation as compared to their counterparts which were grown at room temperature. For example, the as grown film of thickness 380nm had a resistivity of 400 x $10^{-4}$ $\Omega$m, which on ageing reduced to 4 x $10^{-4}$ $\Omega$m, while film of same thickness grown on a heated substrate had a resistivity of 168 x $10^{-4}$ $\Omega$m. However, the resistivity of films grown on heated substrates does not show any systematic variation with thickness. The resistivity of films grown on heated substrates being less than that of as grown films further indicates that ordering contributes in the lowering of resistivity. The XRD carried out on this film continuous to show amorphous nature of the film even after 2 years since deposition. This means both $Sb_2O_3$ (as determined by ESCA) and $Sb_2Te_3$ at the bottom are in amorphous phase as X-ray penetration depth is quite high in diffraction experiments at higher angles. This fact we have shown in fig 8. Therefore, a protective oxide layer prevents $Sb_2Te_3$ to change over to micro-crystalline phase from an amorphous phase. This could well be due to the restriction of degree of freedom for the atomic vibration and hence the atomic re-arrangement leading to micro-crystalline phase ass discussed earlier. Even the diffusion of oxide species into $Sb_2Te_3$ will enable the restriction of $Sb_2Te_3$ rearrangement.

## Conclusion

The ageing of as grown $Sb_2Te_3$ films at room temperature shows a non-linear decrease in resistance due to the transformation of initial amorphous phase to micro-crystalline phase as revealed by X-ray diffraction. This smooth transformation of phases takes place with the atomic rearrangements healing the wrong bonds. In contrast, the ageing of the film grown at an elevated temperature showed an initial linear and a later parabolic increase in film resistance due to surface oxidation resulting from the immediate exposure to air after deposition. The buried $Sb_2Te_3$ layer under the surface $Sb_2O_3$ layer does not show any transformation of amorphous to micro-crystalline phase even after two years since deposition. This could be well due to the restriction imposed on the atomic rearrangements of $Sb_2Te_3$ due to the inter-diffusion of $Sb_2O_3$.

## Acknowledgement

We are grateful to Mr. Padmakshan, Department of Geology, University of Delhi, for carrying out the X-ray diffraction studies.

# Figure Captions

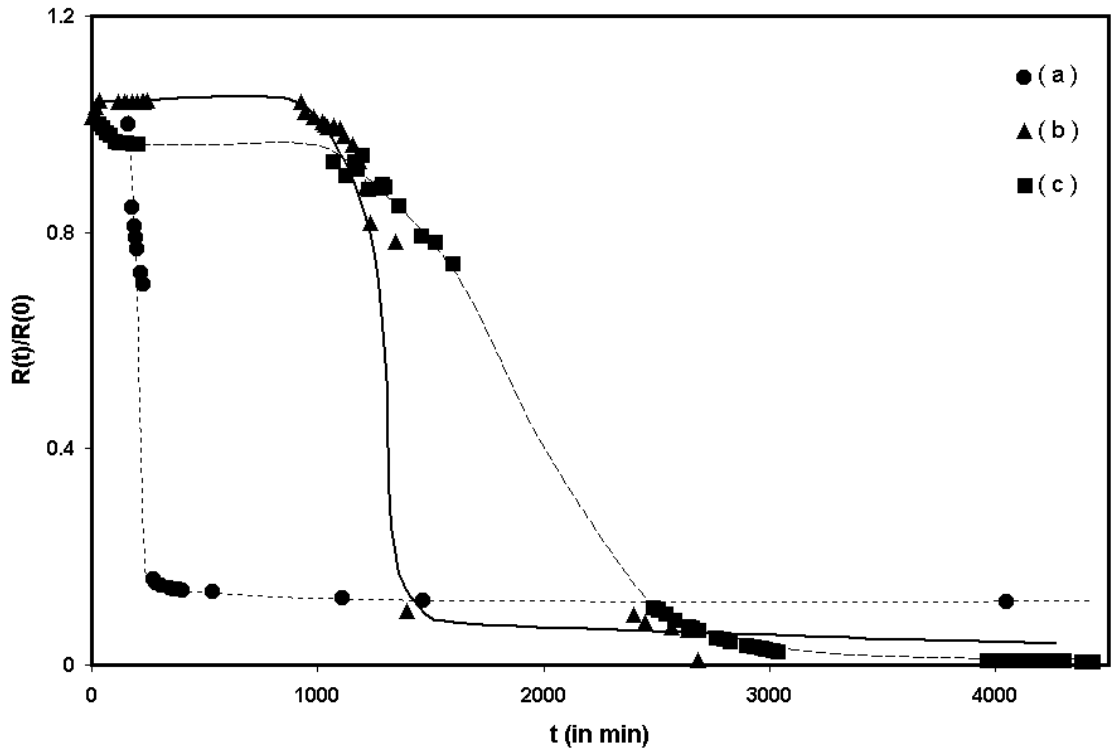

**Figure [1]** The variation in resistance of $Sb_2Te_3$ thin film with passage of time immediately after deposition shown for three different film thickness (a) 500nm (b) 250nm and (c) 160nm.

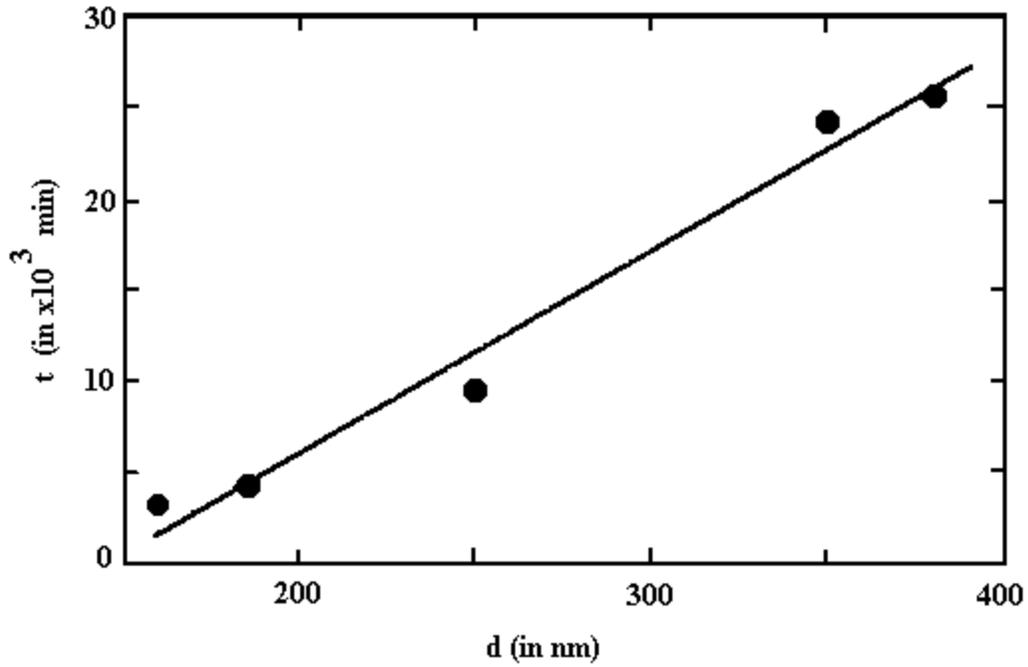

**Figure [2]** The time taken for the resistance of $Sb_2Te_3$ thin film to saturate as a function of film thickness.

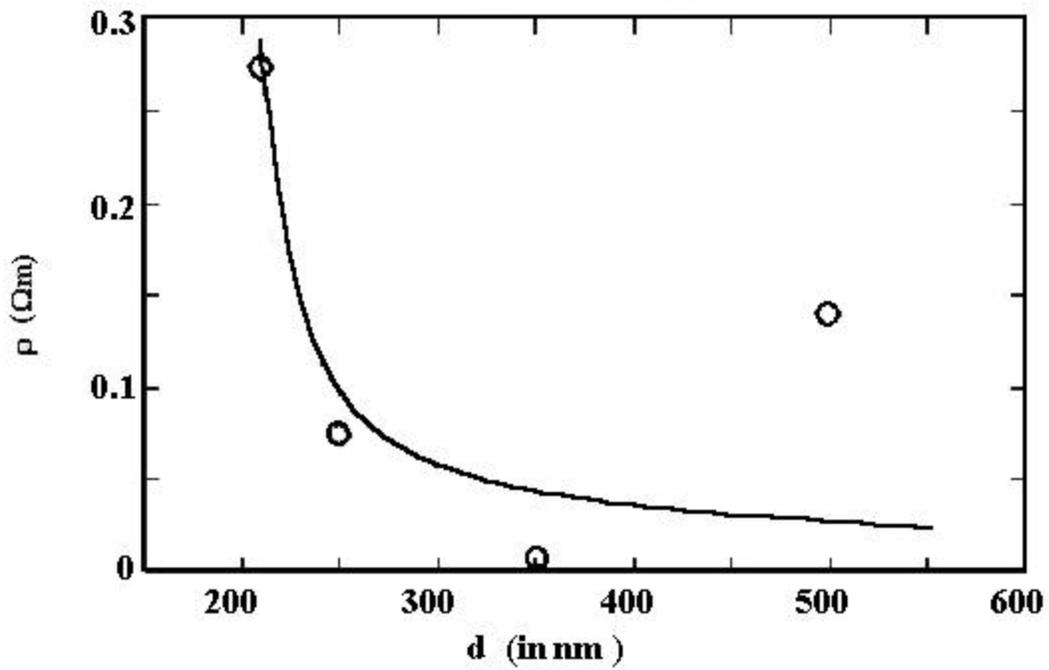

**Figure [3]** The resistivity of as grown film as a function of film thickness.

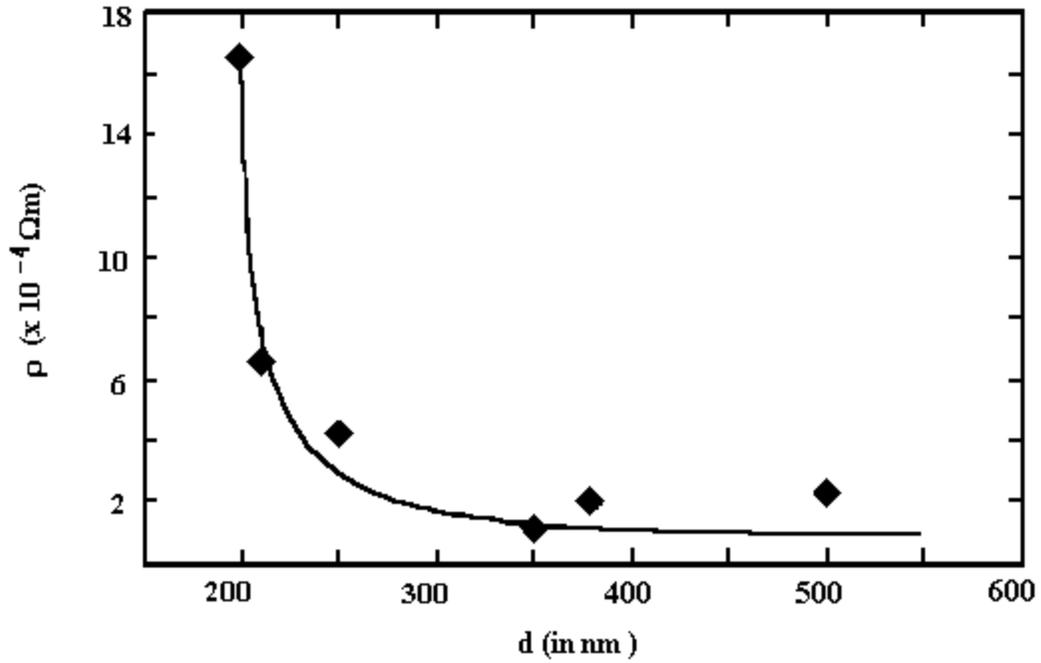

**Figure [4]** The saturated resistivity of as-grown $Sb_2Te_3$ film as a function of film thickness.

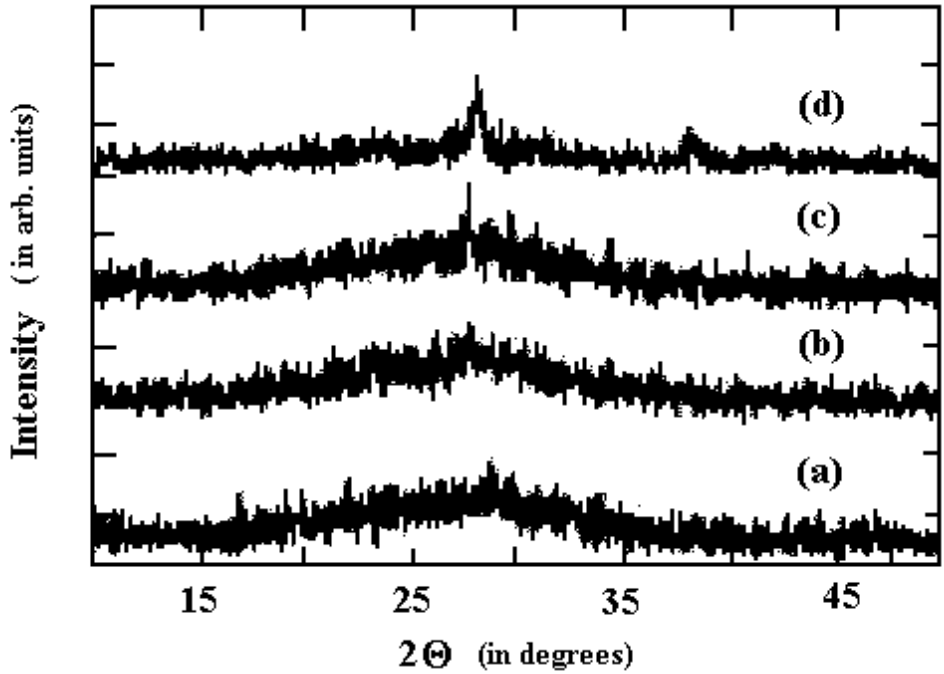

**Figure [5]** X-ray diffractograms of as grown $Sb_2Te_3$ film taken after (a) 40 minutes, (b) 21 hours, (c) 167 hours and (d) 600 hours.

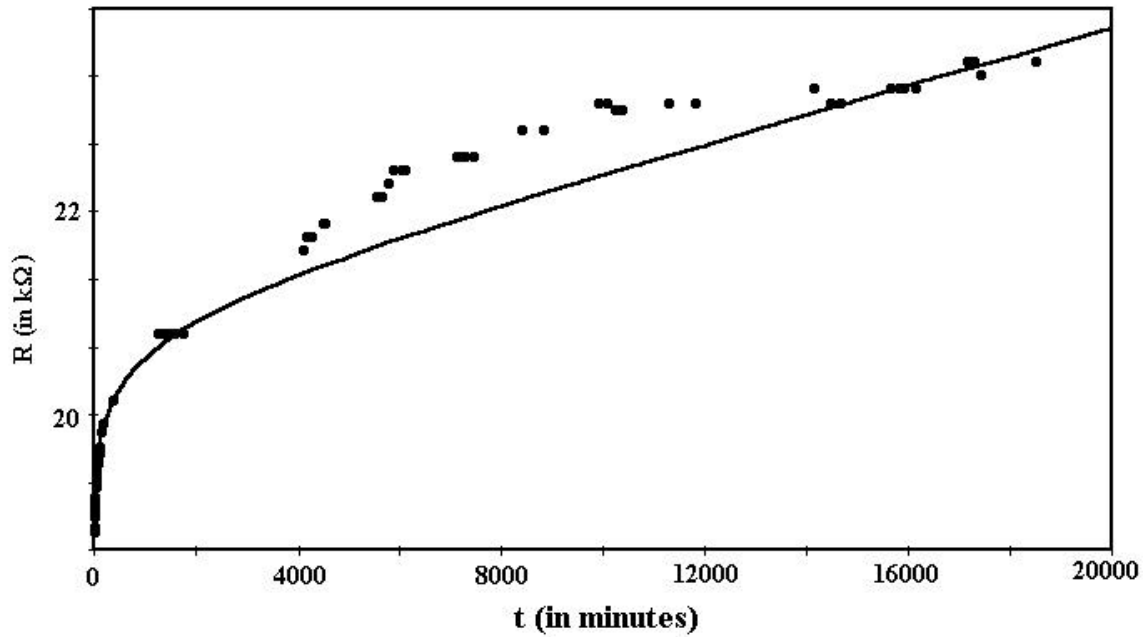

**Figure [6]** The variation in film resistance with time of $Sb_2Te_3$ film grown at substrate temperature of $110^o$C and exposed to air immediately after deposition.

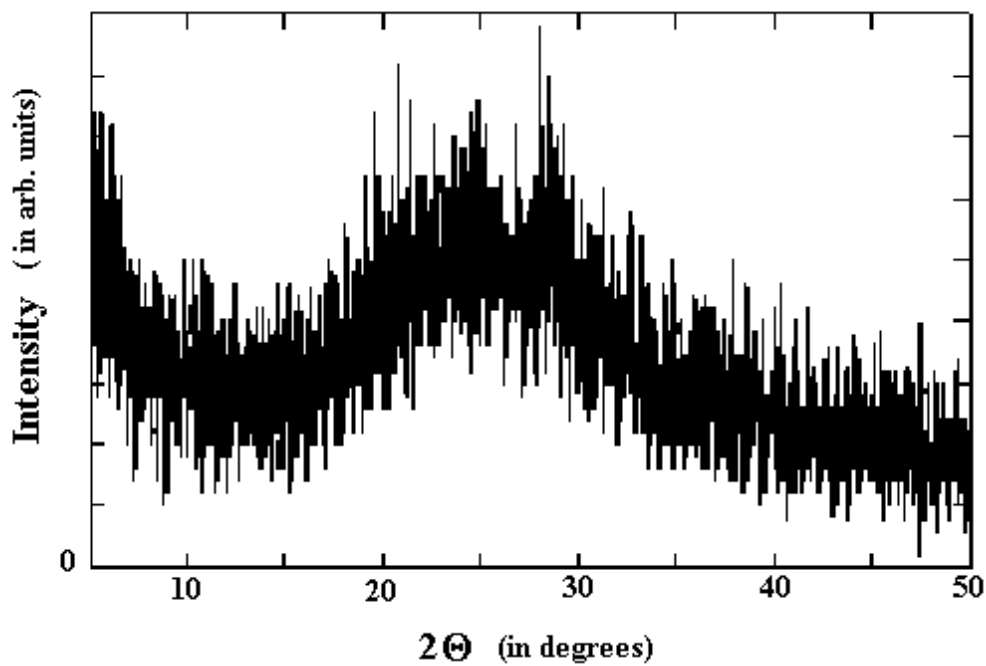

**Figure [7]** ESCA spectra of antimony's 3d peaks.

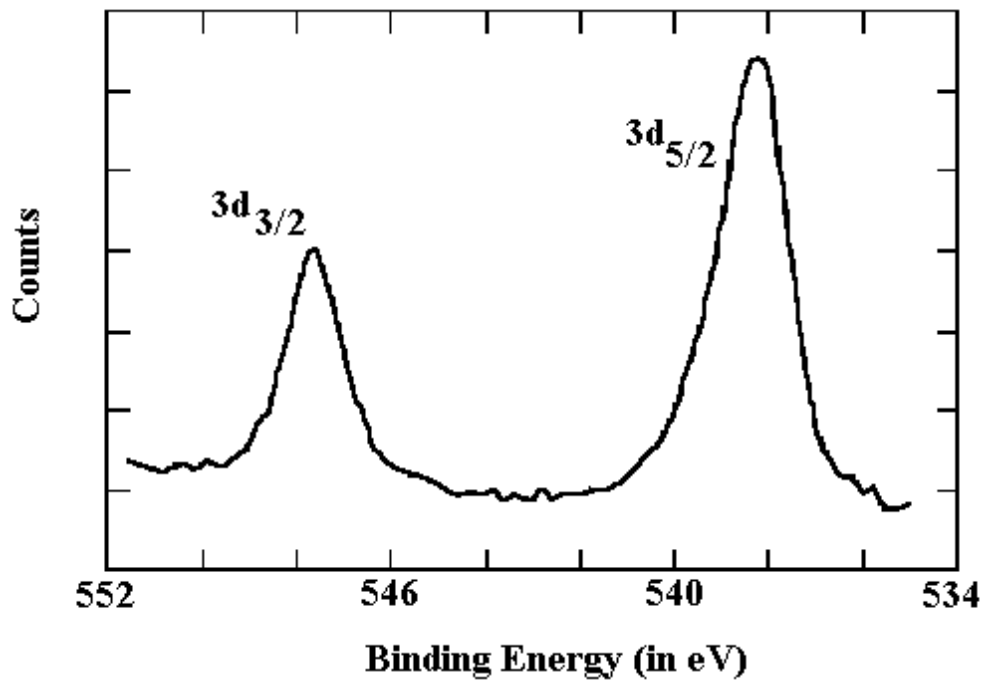

**Figure [8]** X-ray diffractogram of an $Sb_2Te_3$ film of fig 6 taken after 2 years from the date of fabrication.